\RequirePackage{fix-cm}
\documentclass[natbib, smallextended]{svjour3_yk}       

\smartqed  
\usepackage{graphicx}
\usepackage[bookmarks = true, bookmarksnumbered = true, pdfpagemode=None, pdfstartview = FitH, pdfpagelayout = SinglePage, colorlinks = true, urlcolor = blue, citecolor = blue]{hyperref}

\usepackage{lineno}

\journalname{Space Science Reviews}
\begin{document}

\title{Comparison of the deep atmospheric dynamics of Jupiter and Saturn
in light of the Juno and Cassini gravity measurements
}

\subtitle{}

\titlerunning{The deep atmospheres of Jupiter and Saturn}        

\author{Yohai Kaspi         \and
             Eli Galanti    \and 
             Adam P. Showman  \and
             David J. Stevenson   \and
             Tristan Guillot      \and
             Luciano Iess         \and
             Scott J. Bolton
           }


\institute{Y. Kaspi \at
  Corresponding author \\
  Dept. of Earth and Planetary Sciences, Weizmann Institute of
  Science, Rehovot, 76100,  Israel \\
              Tel.: +972-8-9344238\\
              Fax: +972-8-9344124\\
              \email{yohai.kaspi@weizmann.ac.il}           
           \and
           E. Galanti \at
           Dept. of Earth and Planetary Sciences, Weizmann Institute of
  Science, Rehovot, 76100, Israel \\
           \and
            A. P. Showman \at
            Lunar and Planetary Laboratory, University of Arizona,
            Tucson, AZ 85721-0092, USA \\
            \and
            D. J. Stevenson \at
            Division of Geological and Planetary Sciences, California Institute of
            Technology. Pasadena, CA, 91125, USA \\
           \and
           T. Guillot \at
            Universit\'{e} C\^{o}te d'Azur, OCA, Lagrange CNRS, 06304
            Nice, France \\
            \and
            L. Iess \at
            Sapienza Universit\'{a} di Roma, 00184, Rome, Italy \\
            \and
            S. J. Bolton \at
            Southwest Research Institute, San Antonio, TX 78238, USA
}

\date{Submitted: 22-Aug-2019}

\maketitle

\begin{abstract}
The nature and structure of the observed east-west
flows on Jupiter and Saturn has been one of the longest-lasting mysteries
in planetary science. This mystery has been recently unraveled due
to the accurate gravity measurements provided by the Juno mission
to Jupiter and the Grand Finale of the Cassini mission to Saturn.
These two experiments, which coincidentally happened around the same
time, allowed determination of the vertical and meridional profiles
of the zonal flows on both planets. This paper reviews the topic of
zonal jets on the gas giants in light of the new data from these two
experiments. The gravity measurements not only allow the depth of
the jets to be constrained, yielding the inference that the jets extend
roughly 3000 and 9000~km below the observed clouds on Jupiter and
Saturn, respectively, but also provide insights into the mechanisms
controlling these zonal flows. Specifically, for both planets this
depth corresponds to the depth where electrical conductivity is within
an order of magnitude of 1~S~m$^{-1}$, implying that the magnetic
field likely plays a key role in damping the zonal flows.

\keywords{Jupiter \and Saturn \and Juno \and Cassini \and Planetary
  Atmospheres \and Gravity Science }
\end{abstract}

\section{Introduction\label{sec:Introduction}}

\begin{figure*}
\begin{centering}
\includegraphics[scale=0.33]{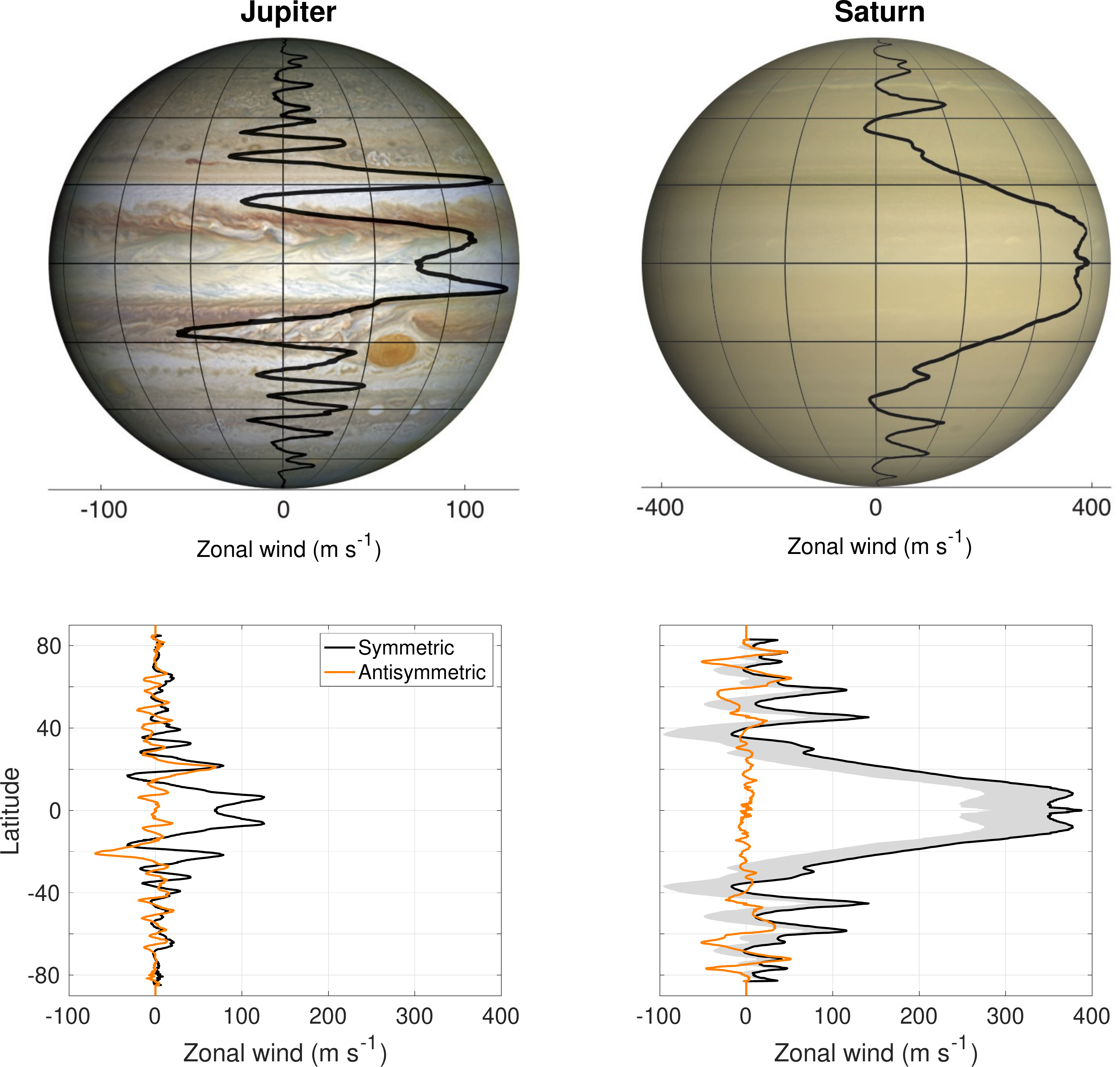}
\par\end{centering}
\caption{\label{fig:Wind} The zonal wind profile of Jupiter (left) and Saturn
(right), divided in the bottom panels into the north-south symmetric
(black) and asymmetric (orange) parts. In the top panels the zonal
wind profile for Jupiter \citep{Tollefson2017} is overlaid on a image
taken by the Hubble Wide Field Camera in 2014, and for the Saturn
case the zonal wind profile \citep{Garcia-Melendo2011} is overlaid
on an image created by Björn Jónsson by combining Cassini and Voyager
images and removing the rings. The grid in both images has a 20$^{\circ}$
latitudinal spread and a $45^{\circ}$ longitudinal spread. The scale
of the zonal flows for Jupiter is the same as the longitudinal grid
on the sphere and for Saturn it is triple.}
\end{figure*}

The most prominent features in the appearance of Jupiter and Saturn
are their east-west banding, which have been observed ever since the
invention of the first telescopes in the 17th century. These so called
'zones' (bright regions) and 'belts' (dark regions) are related to
the two gas giants' east-west jet-streams. The exact interplay between
these zonal flows and the banded structure of the clouds is not completely
understood (see recent review by \citealp{Fletcher2019}), yet the
eastward (westward) jets are typically accompanied by a zone to the
south (north) and a belt to the north (south). Jupiter has about six
distinct jets in each hemisphere (Fig.~\ref{fig:Wind}), including
a wide superrotating eastward jet around the equator and narrower
jets poleward. The strongest jet, reaching $140$~m~s$^{-1}$ is
at latitude $23^{\circ}$N, and is not accompanied by a similar jet
in the southern hemisphere, creating a hemispherical asymmetry (Fig.~\ref{fig:Wind}).
On Saturn, the winds are stronger, with a wider equatorial eastward
flow (up to latitude $\sim30{}^{\circ}$) reaching velocities of nearly
$400$~m~s$^{-1}$. Poleward of that, Saturn has 3-4 distinct jets
in each hemisphere (Fig.~\ref{fig:Wind}). These jet velocities,
measured by cloud tracking (e.g.,~\citealp{Garcia-Melendo2011,Tollefson2017})
and typically quoted relative to Jupiter and Saturn's magnetic field
rotation, have been overall very consistent since the first spacecraft
observations in the 1970s. As Saturn's magnetic field is almost perfectly
axisymmetric, this reference frame has a larger uncertainty for the
case of Saturn, although recent measurements and theoretical calculations
have limited the rotation period uncertainty to within a few minutes
\citep{Read2009,Helled2015,Mankovich2019}.

Prior to the recent Juno and Cassini missions there have been very
little data regarding the flows beneath the cloud tops. The only in-situ
measurements come from the Galileo probe, which descended in 1995
into Jupiter's atmosphere around latitude $6.5^{\circ}$N, and found
that the zonal wind velocity increased from around $80$~m~s$^{-1}$
at the cloud level, where the probe entered, to $\sim160$~m~s$^{-1}$
at a depth of 4~bars, and from there downward the zonal velocity
remained nearly constant down to $21$~bars (130~km) where the probe
was lost \citep{Atkinson1996}. This indicated that the zonal flow
was not restricted to the cloud-level, although the depth where it
was lost is only a mere fraction of the planetary radius, and thus
this measurement did not provide definitive tests to separate between
theories suggesting the flows are a shallow atmospheric phenomena
(e.g.,~\citealp{Williams1978,Williams1979,Cho1996b}), and theories
suggesting the surface flows are just a surface manifestation of deep
cylindrical columns extending deep into the planetary abyss (e.g.,~\citealp{Busse1976,Heimpel2005}).
On Saturn, Cassini observations indicate that low-latitude winds seem
to be stronger at the 2-3 bar level than at the cloud-level (0.5 bar),
while mid-latitude winds seem to be nearly constant or weaker with
depth \citep{Choi2009,Studwell2018}. Above the cloud layer, tracking
wind velocities is difficult as the wind shear can only be indirectly
inferred based on temperature measurements \citep{Simon-Miller2006,Fletcher2007}.
Generally for both planets it seems that winds decay above the cloud-level
but these measurements are very uncertain \citep{Sanchez-Lavega2019}.

The question of how deep the observed jets extend has been debated
extensively in the literature since the early observations by the
Pioneer and Voyager missions in the 1970s. Particularly, the research
has split into two different approaches for how to explain the jets.
According to the first approach, the jets are suggested to be a shallow
atmospheric feature, as appears on a terrestrial planet, thus assuming
all the dynamics are limited to a shallow weather-layer. Geostrophic
turbulence theory provides good understanding to what sets the jet
width and overall number of jets \citep{Rhines1975,Held1996,Chemke2015b},
and matches the numbers observed on Jupiter and Saturn. There have
been many shallow-type models which have showed formation of jets
similar to those on Jupiter and Saturn beginning with the models of
\citet{Williams1978,Williams1979}, and over the years evolved to
more complex models showing formation of multiple jets (e.g.,~\citealp{Panetta1993,Vallis1993,Cho1996,Huang1998,Lee2004,Smith2004,Showman2007,Kaspi2007,Scott2007}).
These shallow-type models most commonly do not exhibit superrotation,
but with particular configurations of bottom drag, internal heating,
moist convection or thermal damping they can produce an equatorial
superrotating jet and multiple high latitude jets \citep{Scott2008,Lian2008,Lian2010,Liu2010,Warneford2014,Young2019,Spiga2020}.
The second approach considers deep convection models, in which the
source of the jets is suggested to be internal convection columns
that interact to form the jets seen at the surface. These ideas have
also emerged in the 1970s with the seminal papers of \citet{Busse1970,Busse1976},
and evolved to more complex interior convention 3D simulations (e.g.,~\citealp{Busse1994,Sun1993,Christensen2001,Aurnou2001,Wicht2002,Heimpel2005,Kaspi2009,Jones2009,Heimpel2016}).
These models naturally exhibit superrotation driven by the convergence
of convectively driven momentum near the equator, but do not naturally
produce the multiple jet structure that appears at the higher latitudes.
These two approaches have been debated greatly over the past several
decades, but due to the lack of observational evidence, the debate
has remained unresolved (see reviews by \citealp{Vasavada2005} and
\citealp{Showman2018}). Now, following the Juno and Cassini gravity
measurements \citep{Iess2018,Iess2019}, which are reviewed here,
the discussion about the source and structure of the jets can be reinvigorated
by these new evidence.

The Juno gravity experiment is one of the key objectives of the Juno
mission \citep{Bolton2005}, with the purpose of measuring Jupiter's
gravity spectra to high accuracy, and thereby providing information
about Jupiter's interior and atmospheric flows \citep{Hubbard1999,Kaspi2010a}.
Juno has been in orbit around Jupiter since July 2016, orbiting Jupiter
every 53 days with X and Ka-band radio links to Earth allowing measurements
of Jupiter's gravity field via Doppler shifts in the radio frequencies
sent to Earth \citep{Bolton2017}. The measurements are obtained around
the time of closest approach (perijove) at 4000~km above the cloud
level \citep{Iess2018}. The perijoves are designed to give an overall
360$^{\circ}$ longitudinal coverage of Jupiter's as the planet rotates
underneath the orbiting spacecraft. In addition, due to the oblateness
of Jupiter the perijoves drift about 1$^{\circ}$ in latitude poleward
every orbit, with the first being at latitude $3^{\circ}$N. As the
Juno microwave radiometer and the Ka-band radio experiment can not
operate in tandem only a subset of the orbits have been devoted to
gravity. Nonetheless, there have been enough gravity orbits to date
that the error estimate of the measured zonal harmonics has reached
saturation.

Motivated by the Juno mission polar orbital configuration, it was
decided that during the Cassini Grand Finale (the final Cassini orbits
before terminating the mission by a decent into Saturn), the spacecraft
would be sent into a polar orbit similar to that of Juno, diving between
the planet and the innermost ring with close, 3500-km flybys \citep{Edgington2016}.
Between May and August 2017, Cassini performed 22 such flybys (every
6 days), out of which six were devoted for gravity science. Similar
to the case of Jupiter, these gravity measurements have allowed the
measurement of Saturn's gravity spectrum up to $J_{10}$, and increased
the accuracy of the known harmonics by more than two orders of magnitude
\citep{Iess2019}.

In light of these two monumental new experiments, this paper provides
a comparative review of what was learned from the gravity measurements
regarding the atmospheric and interior dynamics on Jupiter and Saturn.
In section \ref{sec:Theory} we briefly review the dynamical relations
connecting the momentum and gravity fields. In section \ref{sec:Measurements}
we review the gravity measurements and compare the measured fields
on both Jupiter an Saturn. The interpretation of these results in
terms of the resulting vertical and meridional profile of the zonal
flows that best matches the gravity measurements is shown in section
\ref{sec:Inversion-of-gravity}. In section \ref{sec:Interaction-with-magnetic}
we discuss the implications of the commonalities between the Juno
and Cassini results, and on what these imply about the possible mechanisms
that affect the flow at depth and how the flow might interact with
the magnetic field. In section \ref{sec:Uranus-and-Neptune} we discuss
similar gravity constraints for the zonal flows on Uranus and Neptune,
and we conclude in section \ref{sec:Discussion-and-Conclusion}.

\section{Theory\label{sec:Theory}}

The theoretical starting point for understanding the zonal jet dynamics
are the Euler equations in the rotating frame

\begin{eqnarray}
\left(\mathbf{u\cdot\nabla}\right)\mathbf{u}+2\Omega\times\mathbf{u}+\Omega\times\left(\Omega\times\mathbf{r}\right) & = & -\frac{1}{\rho}\nabla p+\nabla V,\label{eq:equation of motion rotating frame}
\end{eqnarray}
where $\mathbf{u}$ is the 3D velocity vector, $\rho$ is density,
$p$ is pressure and $V$ is the body force. The rotation rate of
Jupiter is given by the System III rotation \citep{Riddle1976,May1979},
with $\Omega=1.75\times10^{-4}$ corresponding to a period of $9.92$
hours. For Saturn there has been significant uncertainty in its rotation
rate due to the axisymmetric nature of the planets's magnetic field,
but recent studies, using gravity measurements, have constrained the
rotation period to $10.57\pm0.03$ hours \citep{Helled2015,Mankovich2019}.
As both giant planets are rapid rotators, for the purpose of studying
the large scale zonal flows, the Rossby number, which is the ratio
of the inertial accelerations (first term on lhs in Eq.~\ref{eq:equation of motion rotating frame})
and the Coriolis accelerations (second term on lhs in Eq.~\ref{eq:equation of motion rotating frame}),
is small. Thus, in the limit of small Rossby number, the fluid is
in geostrophic balance \citep{Pedlosky1987}, meaning:
\begin{figure*}
\begin{centering}
\includegraphics[scale=0.445]{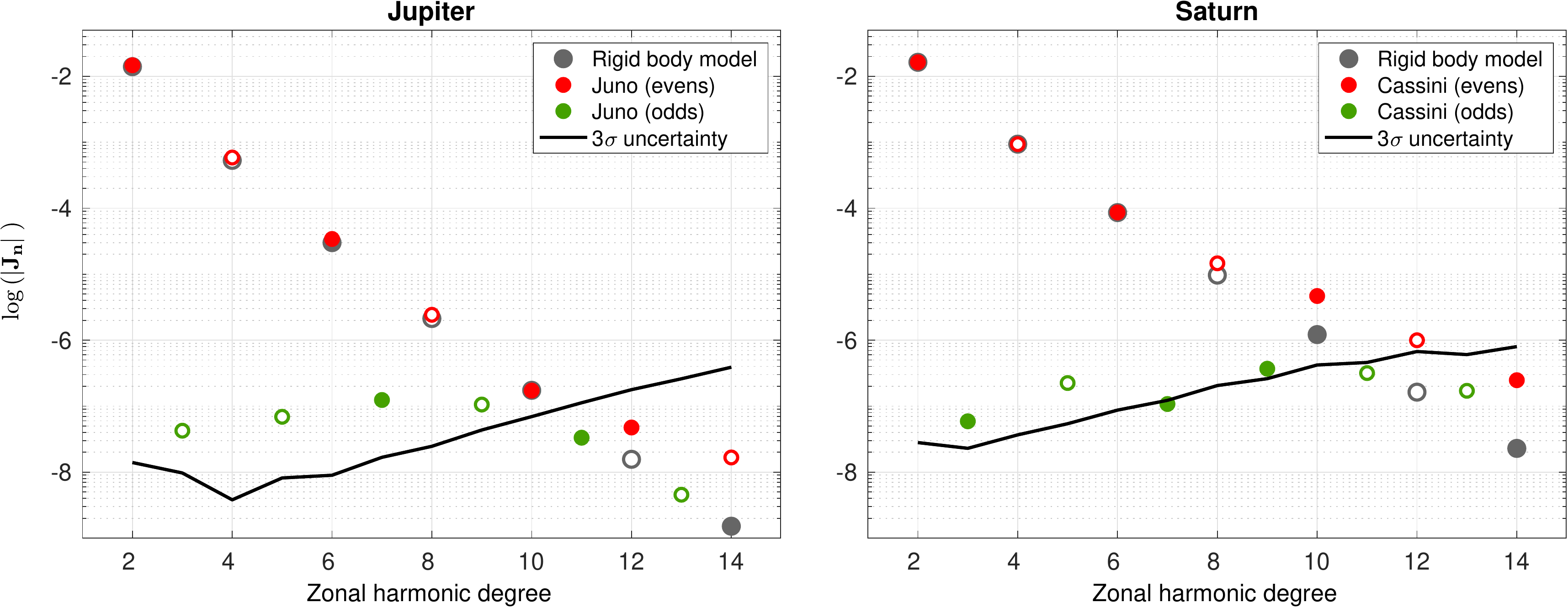}
\par\end{centering}
\caption{\label{fig:Measurememts}The measured zonal gravity harmonics of Jupiter
and Saturn divided between the even harmonics (red) and odd harmonics
(green). For reference the rigid-body harmonics calculated by the
CMS model \citep{Hubbard1999,Hubbard2012} are shown as well (gray).
Full circles denote positive values on the log scale and open circles
denote negative values. The lines are the Juno and Cassini measurement
$3\sigma$ uncertainty \citep{Iess2018}.}
\end{figure*}

\begin{eqnarray}
2\Omega\times\rho\mathbf{u} & = & -\nabla p-\rho\mathbf{g^{*}},\label{eq:geostrophic balance}
\end{eqnarray}
where \textbf{$\mathbf{g^{*}}$} is the effective gravitational field,
$\mathbf{g^{*}}=-\nabla V+\Omega\times\left(\Omega\times\mathbf{r}\right)$,
with the second term being the centrifugal acceleration. Multiplying
equation~\ref{eq:geostrophic balance} by the density $\rho$, and
taking its curl gives

\begin{equation}
2\Omega\cdot\nabla\left(\rho\mathbf{u}\right)=\nabla\rho\times\mathbf{g^{*}},\label{eq: thermal wind general}
\end{equation}
where the left hand side (lhs) has been simplified using mass conservation
$\nabla\cdot\left(\rho\mathbf{u}\right)=0$ and since the rotation
rate vector is constant. This also implies that $\mathbf{g^{*}}$
can be expressed as a scalar potential meaning that $\nabla\times\mathbf{g^{*}}=0$,
which has been used for the rhs of Eq.~\ref{eq: thermal wind general}.
This thermal-wind like relation \citep{Kaspi2009} is different from
the standard thermal-wind used in atmospheric science for a shallow
atmosphere (e.g., \citealp{Vallis2006}), in that the derivatives
on the lhs are in the direction of the spin axis and not in the radial
direction\footnote{Note that $2\Omega\cdot\nabla=2\Omega\frac{\partial}{\partial z}$,
where $z$ is the direction parallel to the spin vector $\left(\Omega\right)$} (the latter is an approximation that holds when the planetary aspect
ratio between the vertical and horizontal scales is small), and the
rhs involves the full density and effective gravity\footnote{Note that the barotropic limit is not simply when the rhs of Eq.~\ref{eq: thermal wind general}
vanishes, but rather when the lhs changes as well, resulting in $2\Omega\cdot\nabla\mathbf{u}-2\Omega\nabla\cdot\mathbf{u}=0$.
See full derivation in \citet{Kaspi2016}.}. Thus, this is a general expression applicable for a rotating atmosphere
at any depth as long as the Rossby number is small.

A considerable simplification to this equation can be taken by assuming
spherical symmetry. Without this assumption the rhs will involve several
terms coming from the deviation of gravity from radial symmetry (both
due to the planetary oblateness and dynamical contributions to the
gravity vector) and the centrifugal terms (see Eq.~\ref{eq:full azimuthal vorticity equation}).
In \citet{Galanti2017a} and \citet{Kaspi2018} a careful treatment
of all these terms is taken, and it is shown that to leading order
Eq.~\ref{eq: thermal wind general} is given by
\begin{equation}
2\Omega\cdot\nabla\left(\rho_{s}\mathbf{u}\right)=\nabla\rho'\mathbf{\times}\mathbf{g_{s}},\label{eq:thermal wind perfect sphericity}
\end{equation}
where $\rho$ has been split into static $\rho_{s}\left(r\right)$
and dynamical $\rho'\left(r,\theta\right)$ components, $r$ is the
radial direction and $\theta$ is latitude. Here $\mathbf{g_{s}}$
is the radial gravitational acceleration coming from integrating $\rho_{s}$.
It is important to note that if the spherical assumption is not taken
in Eq.~\ref{eq: thermal wind general} the rhs evolves into several
different terms of equal magnitude \citep{Galanti2017a}, and using
only part of them \citep{Zhang2015} leads to an inconsistent expansion
(see more detail below). Since the flows on the giant planets are
predominantly in the zonal direction, taking the zonal components
of Eq.~\ref{eq:thermal wind perfect sphericity} allows integrating
the flow induced density gradient to give the dynamical contribution
to the gravity harmonics given by
\begin{equation}
\Delta J_{n}=-\frac{2\pi}{Ma^{n}}\intop_{-1}^{1}d\mu\intop_{0}^{R\left(\mu\right)}r^{n+2}P_{n}\left(\mu\right)\rho'\left(r,\mu\right)dr,\label{eq:delta_Jn}
\end{equation}
where $M$ is the planetary mass, $a$ is the planetary mean radius,
$R$ is the 1-bar radius, $P_{n}$ are the associated Legendre polynomials
and $\mu=\sin\left(\theta\right).$ Note that when integrating $\frac{\partial\rho'}{\partial\theta}$
from the zonal component of Eq.~\ref{eq:thermal wind perfect sphericity},
for use in Eq.~\ref{eq:delta_Jn}, an undetermined radially dependent
integration function arrises $\left(\rho'_{0}\left(r\right)\right)$.
However such a function will not project onto the gravity harmonics
when multiplied by the $P_{n}$ in Eq.~\ref{eq:delta_Jn}, since
\begin{equation}
\intop_{-1}^{1}d\mu\intop_{0}^{a}r^{n+2}P_{n}\left(\mu\right)\rho'_{0}\left(r\right)dr=0,\label{eq:zero_integral}
\end{equation}
because the latitudinally dependent associated Legendre polynomials
$P_{n}$ have a zero mean. Therefore in spherical geometry the dynamical
gravity anomalies can be uniquely determined, despite the density
anomaly itself being determined only up to an unknown constant of
integration \citep{Kaspi2016}.

There has been debate in the literature whether an additional term,
namely $\nabla\rho_{s}\times\mathbf{g'}$ which appears to be of the
same order as the rhs of Eq.~\ref{eq:thermal wind perfect sphericity}
should be included in that equation (termed the thermal-gravity wind
equation by \citealp{Zhang2015}). However, this additional term contains
a deviation from radial symmetry and therefore it was dropped going
from Eq.~\ref{eq: thermal wind general} to Eq.~\ref{eq:thermal wind perfect sphericity}.
If this term is retained, then for consistency, other terms that involve
deviation from radial symmetry, and are of the same order from Eq.~\ref{eq: thermal wind general},
must to be retained as well \citep{Galanti2017a}. Then the azimuthal
component of Eq.~\ref{eq: thermal wind general} will take the form:
\begin{equation}
2\Omega\frac{\partial}{\partial z}\left(\rho_{s}u\right)=\frac{g_{s}^{(r)}}{r}\frac{\partial\rho'}{\partial\theta}-g_{s}^{(\theta)}\frac{\partial\rho'}{\partial r}+\frac{g'^{(r)}}{r}\frac{\partial\rho_{s}}{\partial\theta}-g'^{(\theta)}\frac{\partial\rho_{s}}{\partial r}+\Omega^{2}\left[\frac{\partial\rho'}{\partial\theta}\cos^{2}\theta+\frac{\partial\rho'}{\partial r}r\cos\theta\sin\theta\right],\label{eq:full azimuthal vorticity equation}
\end{equation}
where $u$ is the velocity component in the azimuthal direction, and
the notation $\frac{\partial}{\partial z}\equiv\cos\theta\frac{1}{r}\frac{\partial}{\partial\theta}+\sin\theta\frac{\partial}{\partial r}$
denotes the derivative along the direction of the axis of rotation.
Note that in the radial symmetric limit the rhs reduces to only the
first term on the rhs which is exactly the azimuthal component of
Eq.~\ref{eq:thermal wind perfect sphericity} giving thermal-wind
balance. Eq.~\ref{eq:full azimuthal vorticity equation} is an integro-differential
equation since both the gravity $\mathbf{g_{s}}$ and $\mathbf{g'}$,
are calculated by integrating $\rho_{s}$ and $\rho'$, respectively.
Although this equation can be solved numerically \citep{Galanti2017a},
the additional terms (terms 2-6 on the rhs) are all small and contribute
very little to the gravity solution. The individual contribution of
each of the terms in Eq.~\ref{eq:full azimuthal vorticity equation}
is shown in \citet{Kaspi2018} for the case of Jupiter, demonstrating
that the first term on the rhs is indeed the leading order term. All
other terms in this equation are at least an order of magnitude smaller,
meaning that taking $\mathbf{g}=\mathbf{g}\left(r\right)$ and neglecting
the centrifugal terms gives the leading order solution. \citet{Galanti2017a}
solves the full equation \ref{eq:full azimuthal vorticity equation}
and shows that the resulting gravity harmonics are very close to those
resulting from using thermal wind balance. Other solutions, such as
retaining only the first and third terms on the rhs of Eq.~\ref{eq:full azimuthal vorticity equation}
\citep{Zhang2015,Kong2018}, are thus inconsistent and invalid.

\section{The Juno and Cassini gravity measurements\label{sec:Measurements}}

\begin{table*}
\begin{centering}
\includegraphics[scale=0.91]{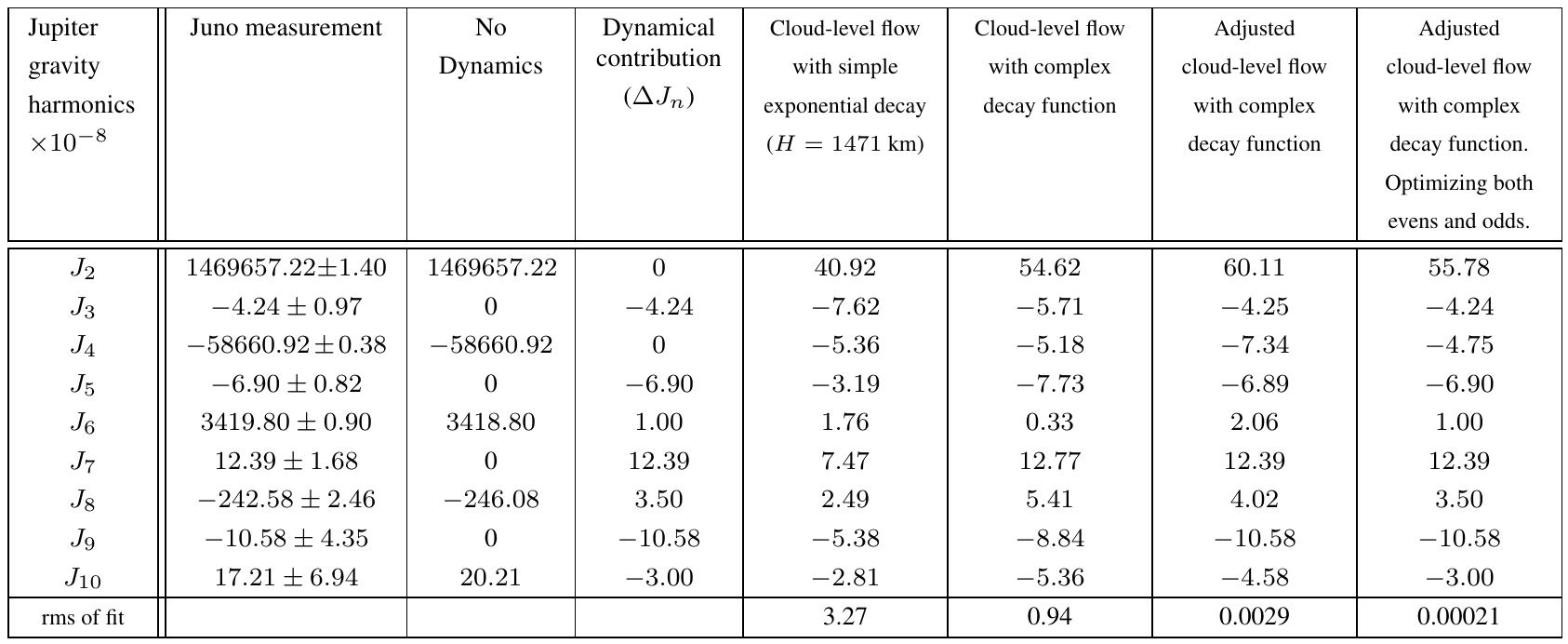}
\par\end{centering}
\caption{\label{tab:Jupiter-values} Values of Jupiter's $J_{2}-J_{10}$ gravity
harmonics. Measured values are the Juno measured values with the uncertainty
being three times the formal uncertainty \citep{Iess2018}. The interior
model values with no dynamics are averages from the ensemble of rigid-body
interior models presented in \citet{Guillot2018}. The dynamical contribution
is the difference between the measured values and the rigid-body models
with no dynamics. The fifth column shows the dynamical solutions when
using the cloud-level profile of the zonal winds (Fig.~\ref{fig:Wind}),
extend down along the direction of the spin axis with a best fit e-folding
decay value of $H=1471$~km (Eq.~\ref{eq: zonal wind exp decay}).
The sixth column shows the solutions when using the cloud level winds,
but with a more complex decay function as described in \citet{Kaspi2018}.
The seventh column shows the solutions when optimizing both the meridional
and vertical profile of zonal wind (as shown in Fig.~\ref{fig:The-vertical-structure},
blue), with the optimization done for the odd gravity harmonics only
(as for columns 5 and 6). The last column shows the same but optimizing
for the even gravity harmonics as well (the profiles are shown in
Fig.~\ref{fig:The-vertical-structure}, green). The last row shows
the normalized rms of the difference between the measurements and
the solution for all cases, giving a relative value for how close
the solution is to the measurements.}
\end{table*}

The close orbits of Juno and Cassini yielded determination of the
gravity harmonics of Jupiter and Saturn to unprecedented accuracy
\citep{Iess2018,Iess2019}. Prior to these missions, the only known
gravity harmonics were $J_{2}$, $J_{4}$ and $J_{6}$ \citep{Jacobson2003,Jacobson2006}.
Supplementary Fig.~1 illustrates how much these have been improved
over the last few decades showing the significant reduction in the
uncertainty going from the Voyager era to the Juno and Cassini measurements.
In addition, the higher-order even harmonics $J_{8}$ and $J_{10}$
have been now determined with high accuracy as well (Tables~\ref{tab:Jupiter-values}
and \ref{tab:Saturn-values}). These even harmonics are mostly affected
by the interior density distribution and shape of the planet, and
only to second order by the flow ($\Delta J_{n}$, Eq.~\ref{eq:delta_Jn}),
although the relative contribution from the flow grows for the higher
harmonics and becomes of similar order to that associated to the rotational
flattening beyond $J_{10}$ \citep{Hubbard1999}. Conversely, the
odd gravity harmonics ($J_{3},$ $J_{5}$, $J_{7}$ etc.) have no
contribution from the interior static density distribution and shape
as these are purely north-south symmetric for such gas planets. The
only possible contribution to the odd gravity harmonics comes from
asymmetries in the dynamics (see the asymmetry in the wind profiles
of both Jupiter and Saturn in Fig.~\ref{fig:Wind}). Therefore, in
terms of probing the dynamics using gravity measurements, the odd
harmonics provide a more direct way of determining the depth of the
flows \citep{Kaspi2013a}.

The values of the even harmonics are to leading order powers of $q^{n}$,
where $q$ is the ratio of the gravity to centrifugal terms in Eq.~\ref{eq:equation of motion rotating frame}
\citep{Hubbard1984}. The rotation therefore is dominant is determining
the values in the rigid-body limit (no dynamics) in addition to the
internal density distribution. The dependence on the density distribution
is more complex, and can be calculated by internal models (e.g.,~\citealp{Hubbard1975b,Hubbard1989,Hubbard1999,Hubbard2012,Nettelmann2012a,Miguel2016,Hubbard2016}),
and depends on the equation of state (EOS) as well (e.g.,~\citealp{Militzer2014,Chabarier2019}).
Both the interior models and the EOS are topics of intense research
and will not be reviewed here, as the focus is on the dynamics. Most
pre Juno/Cassini published interior structure models for Jupiter and
Saturn gave gravity harmonics outside of the narrow range of the Juno
and Cassini measurements (Supplementary Fig.~1), resulting in a need
for improving the interior models and EOSs. The first to match the
Juno measurements to an internal model was \citet{Wahl2017}, who
found that the Juno measured harmonics can only be matched if Jupiter
has a dilute core that expands to a significant fraction of the planet\textquoteright s
radius. These results, however, heavily depend on the choice of EOS.
Recently, \citet{Debras2019} presented a new model for the interior
of giant planets, and were able to match with a new EOS the Juno measurements
as well as the abundance of heavy elements measured by Galileo. Their
results also support the existence of an extended diluted core enriched
by heavy elements.

Despite the dynamical contribution to the even gravity harmonics being
small relative to the rotation and interior mass distribution, due
to the high accuracy of the Juno and Cassini gravity measurements
the dynamical effects on the even harmonics turned to be larger than
the formal measurement uncertainty (supplementary Fig.~1). Therefore,
without any knowledge of how much mass is involved in the flow (i.e.,
how deep the flows are), the dynamical effects can be regarded as
the effective uncertainty of the gravity harmonics \citep{Kaspi2017,Debras2019}.
In Supplementary Fig.~1 we show this effective uncertainty considering
a wide range of possible flow profiles calculated using the method
described in section \ref{sec:Theory} and assuming no a priori knowledge
about the internal dynamics (i.e., before the Juno and Cassini measurements).
However, given the current knowledge about how deep the cloud-level
flows are, and assuming there are no other significant internal flows
that affect the even harmonics, allows more accurate estimates of
the effective uncertainty where the dynamical correction to the measurements
of the even harmonics is already taken into account. In Supplementary
Fig.~1 we show both the effective uncertainty assuming no knowledge
on the dynamical contribution (yellow shading), and the dynamical
contribution given the knowledge of the flow depth (green dots). For
Jupiter, the odd harmonics allowed getting an independent measure
of the flow depth without need to use the even harmonics; while for
Saturn, as discussed in section \ref{sec:Inversion-of-gravity}, the
even harmonics are needed to determine the flow depth which makes
this effective uncertainty more ambiguous.

\begin{table*}
\begin{centering}
\includegraphics[scale=0.91]{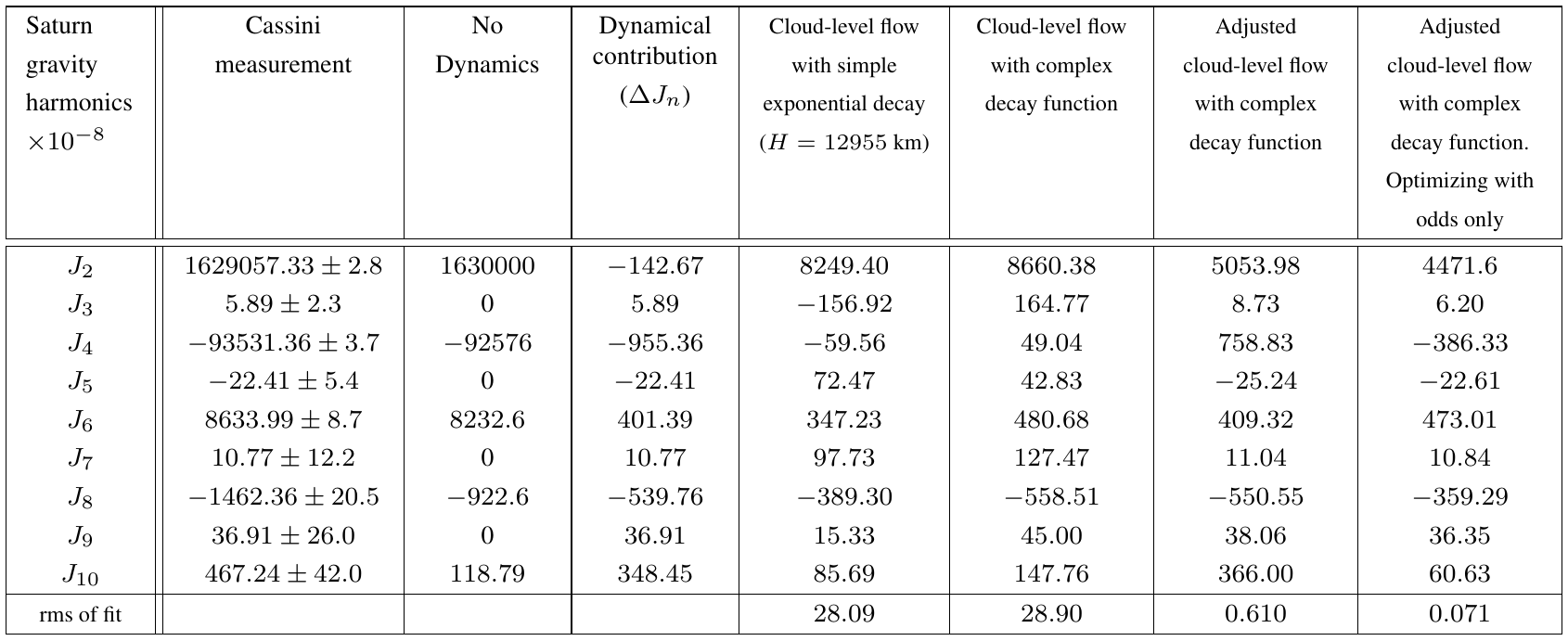}
\par\end{centering}
\caption{\label{tab:Saturn-values} Values of Saturn's $J_{2}-J_{10}$ gravity
harmonics. Measured values are the Cassini Grand Finale measured values
with the uncertainty being three times the formal uncertainty \citep{Iess2019}.
The model values without dynamics are the average values from the
ensemble of models presented in \citet{Galanti2019a}. The dynamical
contribution is the difference between the measured values and the
rigid-body models with no dynamics. The fifth column shows the solutions
when using the cloud-level profile of the zonal winds (Fig.~\ref{fig:Wind}),
and extending them along the direction of the spin axis with a best-fit
e-folding decay value of $H=12995$~km (Eq.~\ref{eq: zonal wind exp decay}).
The sixth column shows the solutions when using the cloud level winds,
but with a more complex decay function as described in \citet{Galanti2019a}.
The seventh column shows solutions when optimizing both the meridional
and vertical profile of zonal wind (as shown in Fig.~\ref{fig:The-vertical-structure},
green) with the optimization done for the even and odd gravity harmonics
together (as for columns 5 and 6). The last column shows the same,
but optimizing for the odd harmonics only (the profiles are shown
in Fig.~\ref{fig:The-vertical-structure}, blue). The last row shows
the normalized standard deviation of the difference between the measurements
solution for all cases, giving a relative value for how close the
solution is to the measurements.}
\end{table*}

The measured even gravity harmonics by Juno and Cassini, as well as
theoretical estimates for the gravity values if Jupiter and Saturn
were rotating as a rigid body (equivalent to a case where the dynamics
are very shallow and have no influence on the gravity field) are presented
in Fig.~\ref{fig:Measurememts} (red and gray dots, respectively).
The numerical values for Jupiter and Saturn are presented as well
in Tables \ref{tab:Jupiter-values} and \ref{tab:Saturn-values},
respectively. As expected, the low order even harmonics match previous
estimates as they are mostly dominated by non-dynamical effects, and
thus very close to the rigid body values. For Jupiter, also $J_{8}$
and $J_{10}$ are relatively close to the rigid-body values and therefore
the measured even harmonics and the rigid-body values are virtually
indistinguishable in Fig.~\ref{fig:Measurememts}. However, for the
Saturn case, these values differ substantially, indicating that the
flows are deeper than on Jupiter. This separation between the measurements
and the rigid-body values matches \citet{Hubbard1999}'s prediction
that if a planet is differentially rotating the even gravity harmonics
beyond $n=8$ will differ from the rigid-body theoretical values.
Hubbard's solution allowed only for cases of full differential rotation,
meaning the surface flows extend throughout the whole planet (i.e.,
following the \citealp{Busse1976} barotropic model). Intermediate
cases for which the surface flows halt at a certain depth can be obtained
using methods as presented in section \ref{sec:Theory} \citep{Kaspi2010a}.
Quantitative solutions showing which vertical decay profiles best
match these measurements are presented in section \ref{sec:Inversion-of-gravity}.

A key result of the gravity measurement was that the measured odd
gravity harmonics vary significantly from zero. The measured values
(green dots in Fig.~\ref{fig:Measurememts}), match predicted theoretical
values \citep{Kaspi2013a} calculated by extending the observed cloud-level
wind inward along the direction of the spin axis and calculating their
affect on the gravity field (see section \ref{sec:Inversion-of-gravity}).
The fact that the expected gravity signal of zonal flows was in fact
detected, confirmed that dynamics play a role in redistributing mass
inside the planet and that the dynamics are deep enough to affect
the measured gravity field. Different from the even gravity harmonics,
the odd harmonics have no contribution from the non-dynamical interior
mass distribution as it should not have any north-south asymmetries
on a gas planet. The observed jets on the other hand do have north-south
asymmetries (Fig.~\ref{fig:Wind}) and are the only considerable
source of north-south asymmetries on the gravity field. Other sources
of north-south asymmetries can be internal oscillations \citep{Durante2017}
and the known north-south asymmetry in the magnetic field \citep{Connerney2018,Moore2018}.
However, internal oscillations can be expected to give fluctuating
contributions from orbit to orbit whereas the measured odd harmonics
are steady. The magnetic effect can be expected to scale as the ratio
of magnetic pressure to total pressure. For a field of $100$~Gauss
(plausibly the unobserved toroidal field) in a region of total pressure
of $\sim100$ kilobars (at $\sim0.96$ Jupiter radii), this is of
order $3\times10^{-9}$, likely too small to be important, although
it cannot be excluded with complete certainty because this field (unlike
the observed poloidal field) is not known. For the case of Jupiter,
$J_{3}$, $J_{5}$, $J_{7}$ and $J_{9}$ were measured to be above
the 3-sigma uncertainty level (black line in Fig.~\ref{fig:Measurememts}),
while for Saturn only the first two are above the 3-sigma uncertainty
level. The robustness of the odd harmonics measurement of Jupiter
allowed therefore to uniquely determine the depth and structure of
the flow even without consideration of the even harmonics. For the
case of Saturn, this turned to be more complex, because only the first
two odd gravity harmonics are above the uncertainty level, and as
shown below those alone do not give a solution that matches the even
harmonics as well.

\section{Inversion of the gravity fields into wind fields\label{sec:Inversion-of-gravity}}

Given the measurements from Juno and Cassini, the challenge is to
translate these measurements into the wind fields that generate them.
The challenge is both in the conversion between the gravity anomaly
data and the dynamically balanced wind field, and in dealing with
the non-unique nature of such solutions. Given that the gravity field
is described by only a finite set of values (Fig.~\ref{fig:Measurememts}),
while a full wind field will require many degrees of freedom to describe
properly, it is obvious that the solution is not unique, and the more
degrees of freedom the wind field has, the easier it will be to find
a fit to the gravity data. We present therefore here a hierarchal
approach beginning with a simple case where the wind is described
with only one degree of freedom, meaning a greater number of observables
(the gravity harmonics), and then present cases with more degrees
of freedom for the wind profile allowing better matches to the gravity
data, but never allow more degrees of freedom for the wind than the
number of overall observables.

\begin{figure*}
\begin{centering}
\includegraphics[scale=0.333]{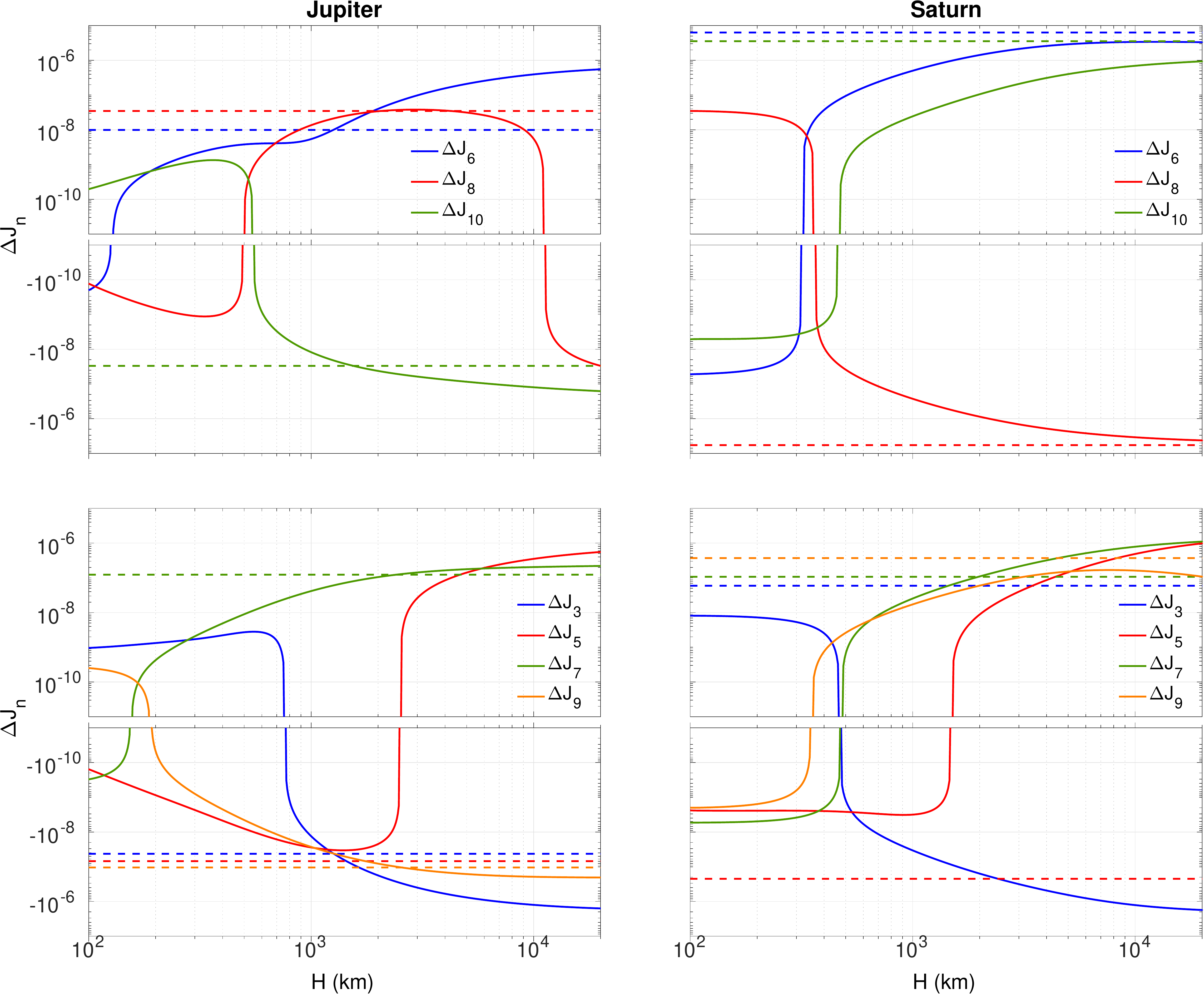}
\par\end{centering}
\caption{\label{fig: predicted_Js}Theoretical values of the even (top) and
odd (bottom) gravity harmonics as function of the e-folding depth
($H$) of the cloud-level wind profile for Jupiter (left) and Saturn
(right). The solid curves are the predicted dynamical contributions
to $J_{n}$ \citep{Kaspi2013a}, for winds decaying exponentially
from the measured cloud-top winds given an e-folding depth $H$. The
horizontal dashed lines are the measured values from Juno and Cassini,
where for the even harmonics the rigid-body values used in Tables
\ref{tab:Jupiter-values} and \ref{tab:Saturn-values} have been
subtracted. Depth values that match the corresponding measured gravity harmonics
correspond to locations where the solid curve crosses the dashed line
of the same color.}
\end{figure*}

We begin with a simple forward model in which we assume the observed
cloud-level flow at the 1~bar level decays radially towards the interior
with an e-folding depth defined as $H$. This represents the expectation
that the wind will overall decay with depth (despite possible enhancement
at the high levels as measured by the Galileo probe), due to the compressibility
of the fluid and/or Ohmic dissipation at depth due to increasing electrical
conductivity. Due to the dominance of rotation, the cloud-level flow
is extended inward along the direction of the spin axis, but the decay
itself is radial since the density growth inward is also radial, meaning
the functional dependence of the zonal flow is given by 
\begin{equation}
u\left(r,\theta\right)=u_{{\rm cyl}}\left(r,\theta\right)\exp\left[\left(r-a\right)/H\right],\label{eq: zonal wind exp decay}
\end{equation}
where $u_{{\rm cyl}}$ is the cloud-level wind profile extended inward
along the direction of the axis of rotation and $H$ is the e-folding
radial decay height of this flow (the other parameters are as defined
in section \ref{sec:Theory}). Such simplified models for the wind
profile have been used in several studies \citep{Kaspi2010a,Kaspi2013a,Liu2013,Kaspi2016,Kong2016,Guillot2018}.

Given such a zonal wind profile, Eq.~\ref{eq:thermal wind perfect sphericity}
can be used to generate the density anomaly gradients that balance
this flow profile, and the dynamical gravity harmonics (Eq.~\ref{eq:delta_Jn})
can be obtained. Fig.~\ref{fig: predicted_Js} (solid lines) shows
such calculated gravity harmonics as function of the e-folding depth
of the flow as predicted in \citet{Kaspi2013a}, for both the even
gravity harmonics (top) and the odd harmonics (bottom). Note that
the values of all harmonics switch sign as function of depth depending
on how the integrated density structure that is balancing the wind
projects on the different spherical harmonics. Although the sign of
these values is not intuitive---the overall tendency to larger values
with depth is---due to having more mass involved with the flow. The
measured values from Juno and Cassini are shown (dashed lines) on
top of the theoretical prediction curves. For the even harmonics,
$\Delta J_{n}$ is calculated as the difference between the measurements
and the average rigid-body values from an ensemble of interior models
(\citealp{Guillot2018,Galanti2019a}; see Tables \ref{tab:Jupiter-values}
and \ref{tab:Saturn-values}).

For the Jupiter case, the measured odd harmonics are all negative
except $J_{7}$ which is positive, matching the prediction for this
simplified model for depths of several thousand kilometers (indicated
by the crossing between the solid and dashed lines in Fig.~\ref{fig: predicted_Js}).
Note that all the four gravity harmonics, independently, match the
\citet{Kaspi2013a} prediction by sign and indicate that the depth
of the flow is between 1000 and 3000 kilometers, with the optimized
best fit e-folding depth for all harmonics combined being $H=1471$~km.
Furthermore, the even gravity harmonics (omitting $J_{2}$ and $J_{4}$
where the relative contribution of the dynamics is very small) show
a similar result where all three theoretical curves cross the measurement
value between depths of 1500 and 2000~km. The fact that for all seven
values ($J_{3}$ and $J_{5}$-$J_{10}$), the theoretical calculation
matches the Juno measurement in sign and value gives a strong indication
that the observed cloud-level flow is related to these measured gravity
anomalies and indicates their depth. Nonetheless, using an exponential
decay law for the cloud-level winds does not give an exact match to
the gravity data (Table~\ref{tab:Jupiter-values}, column 5), and
indeed exponential decay with a uniform e-folding depth was not made
based on physical reasoning but for simplicity. Below we present more
complex decay functions which better match the measurements, yet the
simple model's overall match to the data gives a strong indication
to the relation between the observed flows, their depth and the gravity
measurements.

\begin{figure*}
\begin{centering}
\includegraphics[scale=0.337]{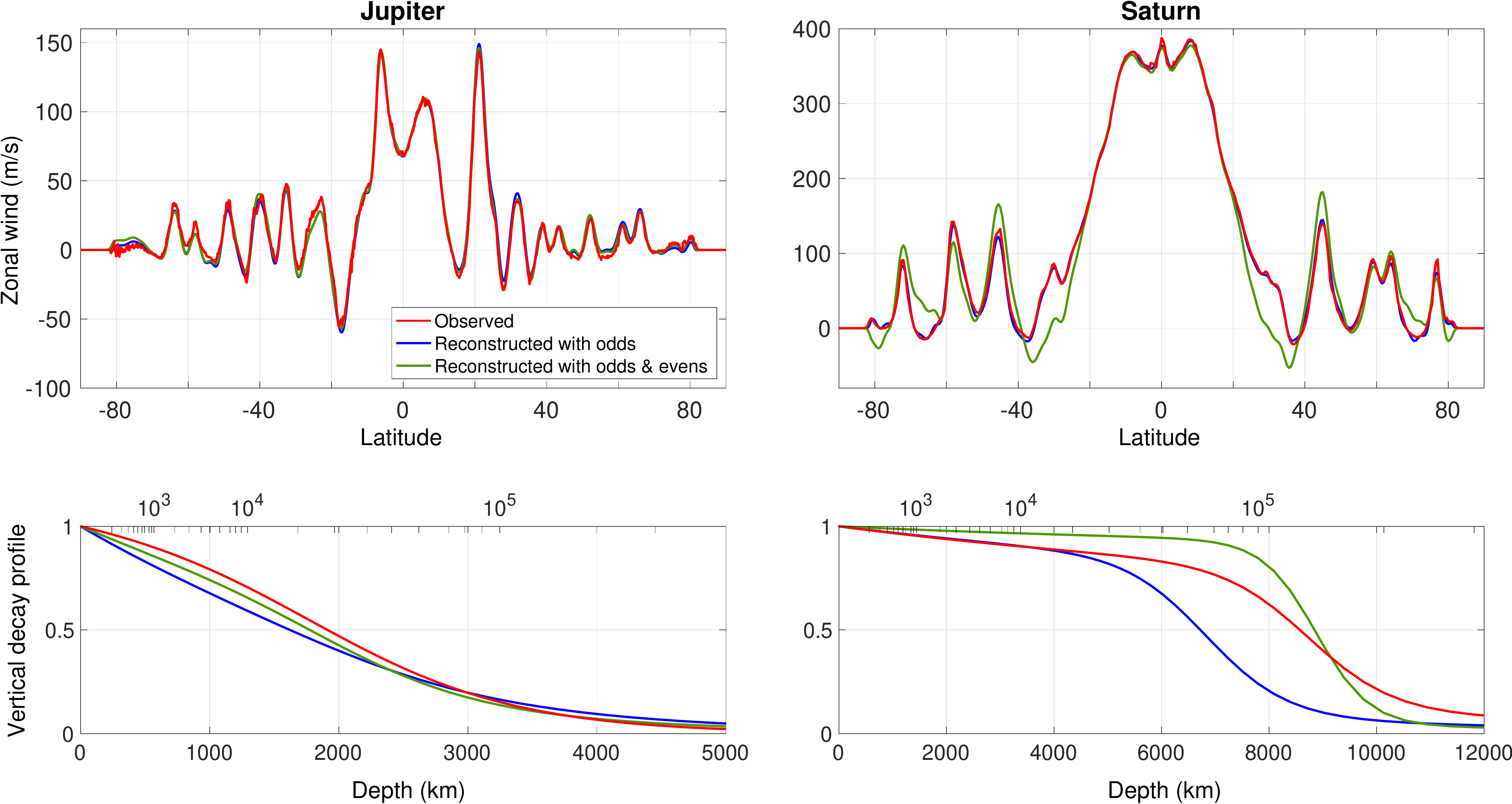}
\par\end{centering}
\caption{\label{fig:optim_profiles}The optimized meridional profile (top)
of the zonal wind for Jupiter (left) and Saturn (right), comparing
the observed cloud-level profile (red), the optimized best fit solutions
taking into account the odd harmonics only (blue), and the optimized
best fit solutions taking into account both the odd and even harmonics
(green). The resulting gravity harmonic values for these profiles
appear in Tables~\ref{tab:Jupiter-values} and \ref{tab:Saturn-values}.
The bottom panel shows the corresponding vertical structures of the
zonal flow as function of depth (km) and pressure (bar).}
\end{figure*}

Optimizing for a more complex decay function we use an adjoint based
inversion technique \citep{Galanti2016}, where a cost function is
minimized to give a best fit between the decay profile and the gravity
measurements, taking into account the uncertainties in the gravity
measurements and the error covariance between the different harmonics
\citep{Kaspi2018}. Solutions for the vertical decay functions using
this method with three degrees of freedom for the shape of the vertical
profile (see \citealp{Kaspi2018} for details) are shown in Fig.~\ref{fig:optim_profiles}.
Taking the exact observed cloud-level zonal flows and extending them
into the interior with this best optimized vertical decay function
gives a much better match to the gravity data than the exponential
decay function (Table~\ref{tab:Jupiter-values}, column 6). Next,
allowing the optimization procedure to include small variations to
the cloud-level wind profile (assuming the zonal wind meridional profile
at depth may vary somewhat from what is observed at the cloud-level)
shows that in this case the solutions give an even better match (Table~\ref{tab:Jupiter-values},
column~7) to all 4 measured odd gravity harmonics (note that even
harmonics are not optimized here, but still give a rather good match).
Fig.~\ref{fig:optim_profiles} shows that in this case the variations
to the observed wind profiles are very minor and well within the uncertainty
(and observed variation between the Voyager and Juno eras) of this
profile. The vertical profile in this case (blue) is very similar
to the one obtained without varying the meridional structure of the
wind profile indicating again the decay being at around several thousand
kilometers.

As the odd harmonics are a consequence of the dynamics alone we have
used only them so far for the optimization procedure. Despite this,
the resulting even $\Delta J_{n}$ for these vertical profiles match
well both in sign and in magnitude the difference between the measurements
of $J_{6}$, $J_{8}$ and $J_{10}$ and the rigid body values (compare
columns 4 and 7 in Table~\ref{tab:Jupiter-values}). Thus it is clear
that if we include the even values in the optimization the results
will not differ substantially. In the final column of Table~\ref{tab:Jupiter-values}
we present such an optimization, where now all values of the seven
gravity harmonics ($J_{3}$ and $J_{5}$-$J_{10}$) match exactly
the measurements. Here again we allow the wind profile to vary from
the observed wind structure, though as can be seen in Fig.~\ref{fig:optim_profiles}
the wind profile needs very minor changes in order to match the gravity
measurements perfectly. We emphasize though that this exercise is
not unique and other meridional and vertical profiles of the zonal
flow can give an exact match to the gravity data. However, following
Occam's razor reasoning, here we have shown that taking the \textit{observed}
cloud-level flow, and extending it inward in a very simple fashion
gives an exact match to the gravity measurements.

Complementary to this analysis, \citet{Guillot2018} used a wide range
of rigid-body models for Jupiter (without averaging as in Table~\ref{tab:Jupiter-values},
column 3) to construct the possible dynamical contribution range to
the even zonal harmonics, by subtracting this range of rigid-body
solutions from the Juno gravity measurements. This range was used
to constrain a wide range of hypothetical flow profiles derived using
thermal wind balance (as in \citet{Kaspi2017}), not bounded to the
observed cloud-level flows with different e-folding decay depths.
This analysis, consistent with the analysis presented above, showed
that with high likelihood the flow extends down to $2000-3500$~km
beneath the cloud-level. Although the \citet{Guillot2018} analysis
is not sensitive to provide the vertical decay profile of the flow,
nor can it constrain the meridional profile of the zonal flow, it
gives an independent method for constraining the bulk depth of the
flow using the even gravity harmonics alone. Given the match of the
flow to the even gravity harmonics it also indicates that the flow
beneath this layer is likely very weak ($<5$~m~s$^{-1}$), otherwise
it would have an influence on the measured even gravity harmonics.

For the Saturn case, the cloud-level flow has shown much more variability
between the Voyager and the Cassini eras \citep{Garcia-Melendo2011},
and is more uncertain. Repeating the same analysis as for the Jupiter
case and taking the cloud-level flow with a simple exponential decay
gives a relatively good match to the even harmonics (same sign and
within factor of 2 in magnitude) for e-folding depths of $\sim10^{4}$
km. This indicates a substantially deeper flow than for the case of
Jupiter. The odd harmonics for the Saturn case, despite being not
very different in magnitude than for Jupiter, are closer to the measurement
uncertainty and therefore only $J_{3}$ and $J_{5}$ have significant
values (Fig.~\ref{fig:Measurememts}). Both, however, give an opposite
sign compared to the theoretical prediction when using the cloud-level
wind profile, indicating that a more sophisticated model is needed
for Saturn. Similarly, as long as using the cloud-level winds, taking
a more complex decay profile does not give a good match to the gravity
measurements (Table~\ref{tab:Saturn-values}, column 6). However,
when allowing the cloud level wind to deviate from the observed profile,
a good match to the measurements can be found (Fig.~\ref{fig:optim_profiles},
green profile).

The deviation from the observed wind profile of Saturn is mainly around
latitude 30$^{\circ}$ where the flow needs to be more westward than
the observed cloud-level wind, with values $\sim50$~m~s$^{-1}$
in order to match the measurements for both the even and odd harmonics
(Table~\ref{tab:Saturn-values}, column 7). The match is obtained
with a vertical profile which is nearly barotropic down to $\sim8000$~km
and then decays with depth (green profile in Fig.~\ref{fig:optim_profiles})
\citep{Galanti2019a}. A similar conclusion, that a westward flow
around latitude 30$^{\circ}$ is needed in order to match the even
gravity harmonics, was also reached by \citet{Militzer2019} who used
a model allowing the flow to extend inward only barotropically (without
changing along the direction of the spin axis), and found that such
a westward zonal flow profile (but twice as large) is needed to match
the measurements. Based on theoretical argument alone, \citet{Chachan2019}
obtained a similar conclusion that a retrograde wind profile is necessary
around latitude 30$^{\circ}$ in order to match the measurements.
The optimizations discussed here used both the values of the odd and
the even harmonics and took into account all cross correlations. For
the case of Saturn, optimizing with odd harmonics only does not give
a good match to the even harmonics, highlighting the difference between
Jupiter and Saturn and pointing to that for Saturn the high order
even harmonics (particularly $J_{8}$ and $J_{10}$) are key to determine
the depth and profile of the deep flows. The uncertainty in rotation
rate affects only the dynamical $J_{2}$ and $J_{4}$ and thus is
not important for interpreting the Saturn gravity measurements \citep{Galanti2017d}.

Comparing the different columns in Tables \ref{tab:Jupiter-values}
and \ref{tab:Saturn-values} and considering the different profiles
in Fig.~\ref{fig:The-vertical-structure} shows that the gravity
results not only inform us about the depth of the jets, but also about
the meridional profile of the zonal flow at depth. The results show
that the measurements are sensitive to the exact meridional profile,
although the variations to it needed to get exact matches are not
significant. To test the statistical significance of this profile,
other profiles with a different meridional profile of the zonal flow
have been tested, to investigate the possibility that the flow at
depth might exhibit major qualitative differences from the flow observed
at cloud-level. Out of a sample of a thousand zonal-wind profiles
as a function of latitude with the same overall amplitude but different
meridional profile, less than 1\% had a better match to the measurements
using the same optimization procedure \citep{Kaspi2018}. These few
profiles had no correlation to one another, nor to the cloud-level
profile. This indicates that although such random solutions can be
found, it is with high confidence that the same meridional profile
of the zonal flow that is observed at the surface extends to depth.

\section{Interaction of the flow with the magnetic field at depth\label{sec:Interaction-with-magnetic}}

\begin{figure*}
\begin{centering}
\includegraphics[scale=0.32]{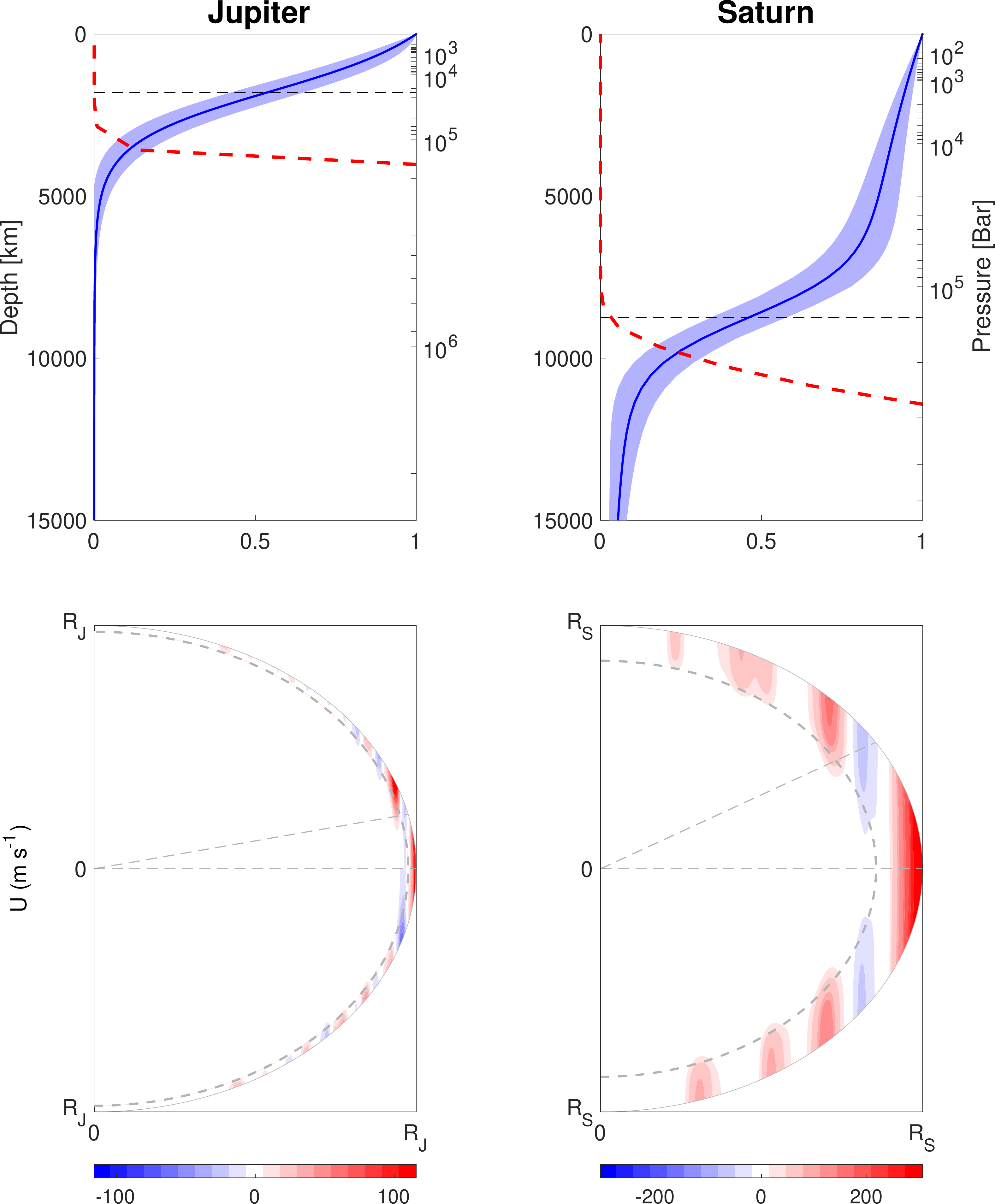}
\par\end{centering}
\caption{\label{fig:The-vertical-structure}The vertical structure of the zonal
flow on Jupiter (left) and Saturn (right) as function of depth corresponding
to the best fit profile presented in column 6 of Table~\ref{tab:Jupiter-values}
and column 7 of Table~\ref{tab:Saturn-values}, respectively. The
upper panel shows this vertical profile function (blue), its uncertainty
(blue shading) and the electrical conductivity profile (red dashed)
as given by \citet{Liu2008} for Jupiter and \citet{French2012} for
Saturn. The electrical conductivity is in units of S~m$^{-1}$ with
the scale going linearly from 0 to 100. The middle point in the decay
profile, at depths of $1831$~km and $8743$~km for Jupiter and
Saturn, respectively, is marked by the dashed horizontal line. The
bottom panels show the same zonal flow profile as function of latitude
and depth in the spherical projection. The middle point corresponding
to that shown in the upper panels appears as the thick dashed line.
The thin dashed lines contain the angle (latitude) derived from extending
the depth of the flow along the direction of the spin axis ($\theta_{e}=\cos^{-1}\left(x/a\right)$,
where $a$ is the planetary radius and $x$ is the depth beyond the
middle point of the flow profile). This latitude ($\theta_{e}$) is
$13^{\circ}$ for Jupiter and $31^{\circ}$ for Saturn, close to the
latitude where the flow is observed to turn from eastward to westward
at the cloud-level.}
\end{figure*}

The results presented above have shown some key similarities and differences
between Jupiter and Saturn. On both planets, the measured gravity
harmonics indicate how deep the cloud-level flows extend, and give
a good match to a zonal wind profile at depth being very similar to
the one observed at the cloud-level of both planets. Differently,
on Jupiter the jets extend down to $\sim3000$ km while on Saturn
that depth nearly triples (Fig.~\ref{fig:The-vertical-structure}).
However, note that the mass of Jupiter is $3.3$ times that of Saturn
while the radius is only about $1.2$ times larger, resulting in the
fact that the gravitational acceleration on Jupiter is about three
times larger than that of Saturn. As a result, the electrical conductivity
of Saturn only achieves large values at a much greater depth than
it does on Jupiter (Fig.~\ref{fig:The-vertical-structure} red dashed
lines in the top panels). Despite the different depth, note that in
both cases the rise in electrical conductivity is near $10^{5}$ bar,
due to the dependence of electrical conductivity on temperature \citep{Stevenson2003}.
Strikingly, the depth where the electrical conductivity rises in both
planets is at the same depth where the gravity measurements imply
that the zonal winds decay (where the blue and red curves cross in
Fig.~\ref{fig:The-vertical-structure}). This strongly hints that
Ohmic dissipation plays a role in damping the flow at depth, and in
fact was predicted previously based on theoretical arguments \citep{Liu2008,Cao2017}.

As a consequence, the depth at which a particular pressure is reached
in Saturn is about three times greater than the corresponding depth
for Jupiter. The temperature at a given pressure is only modestly
($\sim20\%$) lower in Saturn than in Jupiter. Temperature is most
important for determining the electrical conductivity of hydrogen,
which results from excitation of electrons across the band gap between
valence and conduction states \citep{Stevenson1977a}. This conductivity
is an extremely strong function of temperature, both in theoretical
\citep{French2012} and experimental results \citep{Nellis1992},
which are essentially in agreement. The expected conductivity at 3000~km
in Jupiter and 9000~km in Saturn (about $10^{5}$ bars in both planets)
is approximately 1~S~m$^{-1}$ (similar to that of salty water at
room temperature). At this conductivity, strong zonal winds would
create a toroidal magnetic field whose associated electrical currents
would produce a total Ohmic dissipation that is comparable to the
observed luminosity of the planets \citep{Liu2006,Liu2008}.

The factor of three difference in zonal wind depth between Jupiter
and Saturn, together with a remarkable correspondence to the theoretical
argument of \citep{Liu2008} (their prediction was $2800$~km for
Jupiter) strongly suggests the role of magnetohydrodynamics. It should
also be noted that because the electrical conductivity is such an
extremely strong function of temperature and therefore radius, the
results hold even given a likely order of magnitude uncertainty in
the electrical conductivity and the large difference in field strengths
between Jupiter and Saturn. Some cautionary comments are in order,
however: first, the argument is purely kinematic; that is, there is
as yet no fully dynamical argument that explains this truncation of
the flows by the magnetic field. Moreover, the Lorentz force is not
sufficient by itself to dampen the flows from large values (tens of
m~s$^{-1}$) to zero \citep{Cao2017}. Clearly the role of the magnetic
field is more complicated and the full solution to the zonal flow
requires an understanding of the spatial structure of the deviation
from constant entropy throughout the envelope (Eq.~\ref{eq:wind_shear}).

\begin{figure*}
\begin{centering}
\includegraphics[scale=0.32]{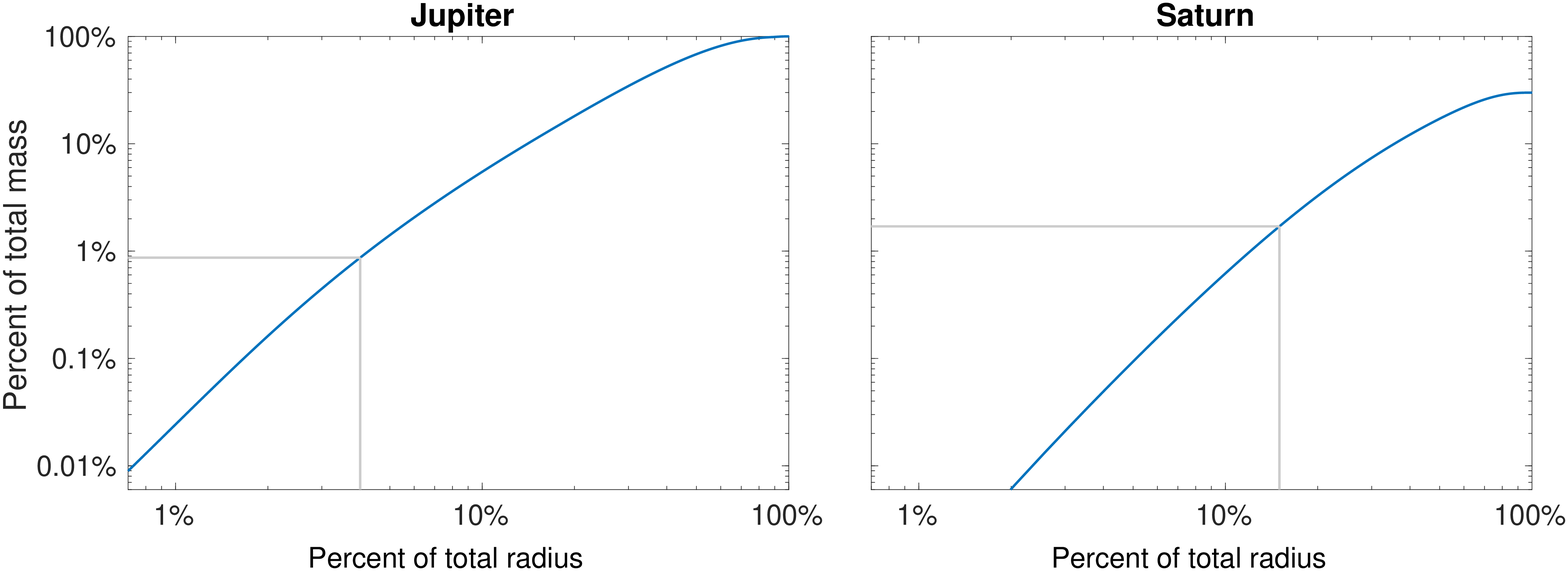}
\par\end{centering}
\caption{\label{fig:mass_frac}The percentage of Jupiter (left) and Saturn's
(right) mass as a function of depth beneath the 1-bar level. The grey
line shows that percentage of mass contained within the depth of the
zonal flows as found by the gravity measurements.}
\end{figure*}

Therefore, although the Ohmic dissipation might be ultimately what
halts the flow at depth, it does not explain what diminishes the strength
of the flow substantially from the cloud-level down to where the electrical
conductivity becomes large (Fig.~\ref{fig:The-vertical-structure}).
In this region the dissipation of the zonal flow needs to be due to
other mechanisms. Reorganization of Eq.~\ref{eq:thermal wind perfect sphericity},
looking at its zonal component and assuming the background profile
is adiabatic leads to a relation between the wind shear along the
direction of the spin axis and the entropy gradients \citep{Kaspi2009}:
\begin{eqnarray}
\frac{\partial u}{\partial z} & = & -\frac{g}{2\Omega\rho_{s}}\left(\frac{\partial\rho}{\partial s}\right)_{p}\frac{1}{r}\frac{\partial s'}{\partial\theta}\label{eq:wind_shear}
\end{eqnarray}
where $s$ is entropy. Thus the direction of the shear is determined
by the average sign of the entropy gradients, and the shear is expected
to be largest at the lower depths where the density is smallest. Note
that in the purely barotropic limit, the rhs of equation vanishes
leading to the flow being purely aligned with the axis of rotation.
This is similar to the Taylor-Proudman theorem \citep{Pedlosky1987},
only that the Taylor-Proudman theorem requires the fluid to be incompressible,
in which case all three components of velocity are aligned with the
rotation axis ($2\Omega\cdot\nabla\mathbf{u}=0$ ). For a compressible
flow, in the barotropic limit, the alignment is for the zonal and
meridional component of the flow (throughout this paper only the zonal
component of the velocity is discussed). Yet, due to the entropy gradients
(Eq.~\ref{eq:wind_shear}), and as evident from the gravity results,
the flow is likely not fully barotropic (baroclinic). Quantitatively
however, the shear on both planets is of O(100 m~s$^{-1}$) over
thousands of kilometers, implying that the flows in effect are not
very far from barotropic. If, for example, we assume the variation
from barotropic is very large and allow the observed flows to extend
inward in a different way (e.g., radially) the match to the gravity
measurements would not have been as good as it is here (Tables~\ref{tab:Jupiter-values},
\ref{tab:Saturn-values}). For example, extending the cloud-level
flow inward radially for Jupiter results in $J_{3}$ changing sign.
Only if the zonal flow meridional profile is altered substantially
solutions can be found.

An additional question regarding the zonal flows is the issue of their
forcing mechanism. It is well understood from terrestrial atmospheric
dynamics that geostrophic turbulence on a rotating planet will drive
turbulent eddy momentum fluxes resulting in regions of momentum flux
convergence with eastward (prograde) flows and momentum flux divergence
with westward (retrograde) flows. It has been observed for Jupiter
\citep{Salyk2006} and Saturn \citep{DelGenio2007} that indeed there
is a strong correlation between the regions of eddy momentum flux
convergence (divergence) and the eastward (westward) jets, implying
this is a plausible mechanism for driving the zonal flows. Yet, it
is not known what is the source of the eddies, with candidates being
barotropic or baroclinic instabilities (e.g.,~\citealp{Kaspi2007})
or the internal convection itself. It has also been shown that both
shallow and deep forcing can drive such zonal flows \citep{Showman2006}.
The driving and dissipation mechanisms discussed here can be seen
in a single expression by taking the leading component of the momentum
equation (Eq.~\ref{eq:equation of motion rotating frame}) and expressing
it in terms of angular momentum \citep{Vallis2006}. Taking a zonal
and vertical average and in steady state gives, $\mathbf{u}\cdot\nabla M=-S+D$
, where $M$ is angular momentum, $S$ is the eddy momentum flux divergence
and $D$ is the Lorentz drag \citep{Schneider2009}. In regions of
low electrical conductivity, closer to the cloud-level, the balance
will be $\mathbf{u}\cdot\nabla M=-S$ with the momentum flux providing
the cross angular momentum surfaces (i.e., across the direction of
the spin axis) flux to force eastward momentum flux convergence and
zonal jets. In the the deep region, where the electrical conductivity
is high, the drag allows for cross angular momentum flow, $\mathbf{u}\cdot\nabla M=D$,
to close the circulation. In between, $\mathbf{u}\cdot\nabla M=0$,
meaning there is no cross angular momentum flow and the flow must
be aligned with the direction of axis of rotation. Note though that
this has no implication on how barotropic is the zonal flow, and this
argument puts a strong constraint on the meridional circulation but
not on the zonal flow itself. These arguments are similar to those
used to explain the midlatitude Ferrel cell on Earth with surface
drag taking the place of the Lorentz drag \citep{Vallis2006}.

The dynamical constraints discussed earlier suggest that the flow
likely extends inward along the direction of the spin axis as demonstrated
in Fig.~\ref{fig:The-vertical-structure} where the full solution
for the zonal wind in the radial-latitudinal plane is presented. It
is evident that for the case of Jupiter, despite the jets being deep
from an atmospheric perspective, extending down to 10$^{5}$ bars
and advecting 1\% of the mass of the whole planet (Fig.~\ref{fig:mass_frac}),
from the point of view of the whole gaseous planet, and compared to
the proposed \citet{Busse1976} model scenario, the winds penetrate
only a small fraction of the planet. For Saturn, the fraction is larger,
going down to 15\% of the radius of the planet, but still containing
only a few percent of the total mass (Fig.~\ref{fig:mass_frac}).
That said, on both planets, the atmospheric advection of several percent
of the planetary mass is very significant, and by far extends the
advection on every other planet in the solar system (e.g., Earth's
atmosphere is less than one part in a million of the planetary mass).

Note that the depth for both Jupiter and Saturn obtained from the
gravity measurements is also consistent with the observed latitudinal
extent of the equatorial eastward flow; meaning that if the depth
of the flow at the equator is extended along the direction of the
axis of rotation, it interests the surface almost exactly at the latitude
where the zonal flow turns from positive to negative (eastward to
westward). Quantitatively, taking the half point of the flow depth
(horizontal lines in the top panels of Fig.~\ref{fig:The-vertical-structure}),
being $1831$~km for Jupiter and $8743$~km for Saturn, and calculating
the latitude where this line intersects the surface gives a latitude
of $13^{\circ}$ for Jupiter and $31^{\circ}$ for Saturn. This is
very close to the latitude where the equatorial flow changes sign
which is $13^{\circ}$ and $35^{\circ}$ for Jupiter and Saturn, respectively.
Note that for the Saturn case if a rotation rate of 10:34 is taken
instead of 10:39 as recent publications indicate this latitude changes
to $31^{\circ}$ (Fig.~\ref{fig:Wind}). Thus, this gives another
independent observation which is in agreement with the conclusion
from the gravity measurements regarding the depth of the zonal jets.
Such arguments regarding the depth of the equatorial superrotation
region have been presented in the past in the context of the tangent
cylinder surrounding the inner region of deep convective models (e.g.,~\citealp{Heimpel2005,Aurnou2008,Kaspi2009,Liu2010,Gastine2013}).

\section{The depth of the zonal flow on Uranus and Neptune\label{sec:Uranus-and-Neptune}}

As on Jupiter and Saturn, Uranus and Neptune also have very strong
east-west flows at the observed cloud-level. These flows reach $\sim200$
m~s$^{-1}$ on Uranus and nearly $400$ m~s$^{-1}$ on Neptune,
with a meridional structure which is overall similar between the two
planets, consisting of a westward broad equatorial flow and a strong
and broad eastward flow at midlatitudes. The flows have an overall
similar character despite the obliquity being very different ($98^{\circ}$
on Uranus and $29^{\circ}$ on Neptune), and the internal heat flux
being three times stronger than the solar flux on Neptune while on
Uranus the internal heat flux appears to be negligible \citep{Pearl1990,Pearl1991}.
As these are the only two planets yet to host a dedicated space mission
\citep{Fletcher2019b}, most data comes from the Voyager encounters
of the two planets in 1986 and 1989 \citep{Smith1986,Smith1989}.
The data obtained from Voyager includes the gravity harmonics up to
$J_{4}$, although to a much lesser precision than the Juno and Cassini
data discussed above \citep{Jacobson2007,Jacobson2009}. Nonetheless,
due to the broader shape of the wind structure and its relative resemblance
to the meridional structure of $P_{4}$ (Eq.~\ref{eq:delta_Jn}),
it was possible to place an upper bound on the depth of the atmospheric
circulation on these planets \citep{Kaspi2013c}.

This was done utilizing the known values of $J_{4}$ from Voyager
and determining the difference between the observed $J_{4}$, and
the $J_{4}$ resulting from a wide range of rigid-body models set
to match all other observational constraints besides $J_{4}$. Any
difference in these quantities places constraints on the dynamical
contribution to $J_{4}$. Therefore, considering the observed $J_{4}$
and its uncertainty and the widest possible range of $J_{4}$ solutions
from interior models, ranging from models with no solid cores to ones
with massive solid cores \citep{Helled2010a,Helled2011,Nettelmann2013},
an upper limit to the dynamic contribution to $J_{4}$ ($\Delta J_{4}$,
as in Eq.~\ref{eq:delta_Jn}) was constructed. This revealed that
the dynamics are constrained to the outermost 0.4\% of the mass on
Uranus and 0.2\% on Neptune, providing a much stronger limitation
to the depth of the dynamical atmosphere than previously suggested
\citep{Hubbard1991}. This result implies that the dynamics must be
confined to a thin weather layer of no more than 1600~km on Uranus
and 1000~km on Neptune \citep{Kaspi2013c}. This is much shallower
than the depths on Jupiter and Saturn with the pressure to which the
flows extends being at most 4000~bar on Uranus and 2000~bar on Neptune.

\section{Discussion and Conclusion\label{sec:Discussion-and-Conclusion}}

The recent gravity measurements of Juno and Cassini have provided
data accurate enough to allow inferring the effect of atmospheric
dynamics on the gravity field of both planets. This allowed determining
the depth of the flows on both planets, and inferring that the meridional
profile of the zonal flows observed at the cloud-level of both planets
likely extends to depth. An intrinsic problem of any gravity inversion
is that the contributing field is not unique, meaning in this case
that the wind-induced gravity anomalies can not be traced uniquely
to the wind field creating them. However, there are several independent
lines of evidence supporting the solutions presented here:
\begin{enumerate}
\item The simplest possible Jovian flow model, taking the observed cloud-level
winds and extending them inward, matches all 4 measured odd gravity
harmonics ($J_{3}$, $J_{5}$, $J_{7}$ and $J_{9}$) independently
both in sign and in magnitude (Fig.~\ref{fig: predicted_Js}).
\item The same wind profile matches the dynamical component of the even
gravity harmonics ($J_{6}$, $J_{8}$ and $J_{10}$) as well (Fig.~\ref{fig: predicted_Js}).
\item There are no other likely sources of north-south asymmetries that
can match in magnitude the measured values of the odd gravity harmonics
(section~\ref{sec:Interaction-with-magnetic}).
\item A statistical analysis taking random zonal wind profiles (considering
that the cloud-level winds may be decoupled from the interior flow
causing the measured gravity anomalies), shows that less than 1\%
of such wind profiles give a match to the gravity measurements \citep{Kaspi2018}.
\item The Saturnian winds, with slight modifications (within the error range
of the wind measurements), match the dynamical component of the gravity
measurements for both the even and odd harmonics (Fig.~\ref{fig:optim_profiles}).
\item For both Jupiter and Saturn, the depth of the flows inferred from
the gravity measurements matches the depth where electrical conductivity
rises abruptly. This suggests that the previously suggested mechanism
of Ohmic dissipation might play a key role in setting the flow depth
\citep{Liu2008}.
\item For both Jupiter and Saturn, the depth of the flows inferred from
the gravity measurements matches the depth inferred from a tangent
cylinder separating the equatorial eastward flow and the higher latitude
flows. This may explain the different latitudinal extent of the equatorial
flow on both planets (Fig.~\ref{fig:The-vertical-structure}).
\item Temporal variation of the magnetic field of Jupiter implies that the
variation is carried by the zonal flow and gives a magnitude of the
zonal flow at depth consistent with the depth implied by the gravity
measurements \citep{Moore2019}.
\end{enumerate}
Overall, although each one of these evidence separately can be perhaps
challenged as being coincidental, when taken together, these consistent
lines of evidence combined yield a coherent picture regarding the
extent and character of the flows beneath the cloud-level of Jupiter
and Saturn. On both planets, the flows advect a substantial part of
the mass of the planets ($1-2\%$), greater than in any other planetary
atmospheres (Fig.~\ref{fig:mass_frac}). This depth implies that
the flows span from top to bottom, nearly four orders of magnitude
in density and nearly six orders of magnitude in pressure. That said,
the depth on both planets is just a small fraction of the planetary
radius ($\sim4\%$ on Jupiter and $\sim15\%$ on Saturn), suggesting
that from a planetary perspective the flows are still bound to a relatively
shallow layer.

Despite this new understanding regarding the depth and structure of
the flow, we are still left with an incomplete picture of the mechanisms
driving the flow. Particularly, several open questions remain, such
as what causes the flow to decay before reaching the Ohmic dissipation
level at $\sim10^{5}$ bar? What drives the equatorial superrotation?
Why are the flows on Saturn substantially stronger? What are the source
of the eddies driving the jets? What are the roles of baroclinic,
barotropic and convective instabilities in driving the winds? With
better constrains on the dynamical component of the gravity fields
(due to improved interior models), magnetic fields and their secular
variation (future Juno orbits), temperature fields and water abundance
(Juno microwave measurements) and improved dynamical models these
questions might be addressed in the coming years, to give a better
understanding of the fundamental physical processes driving the dynamics
on the giant planets.


%
%


\begin{figure*}[b]
\begin{centering}
\includegraphics[scale=0.044]{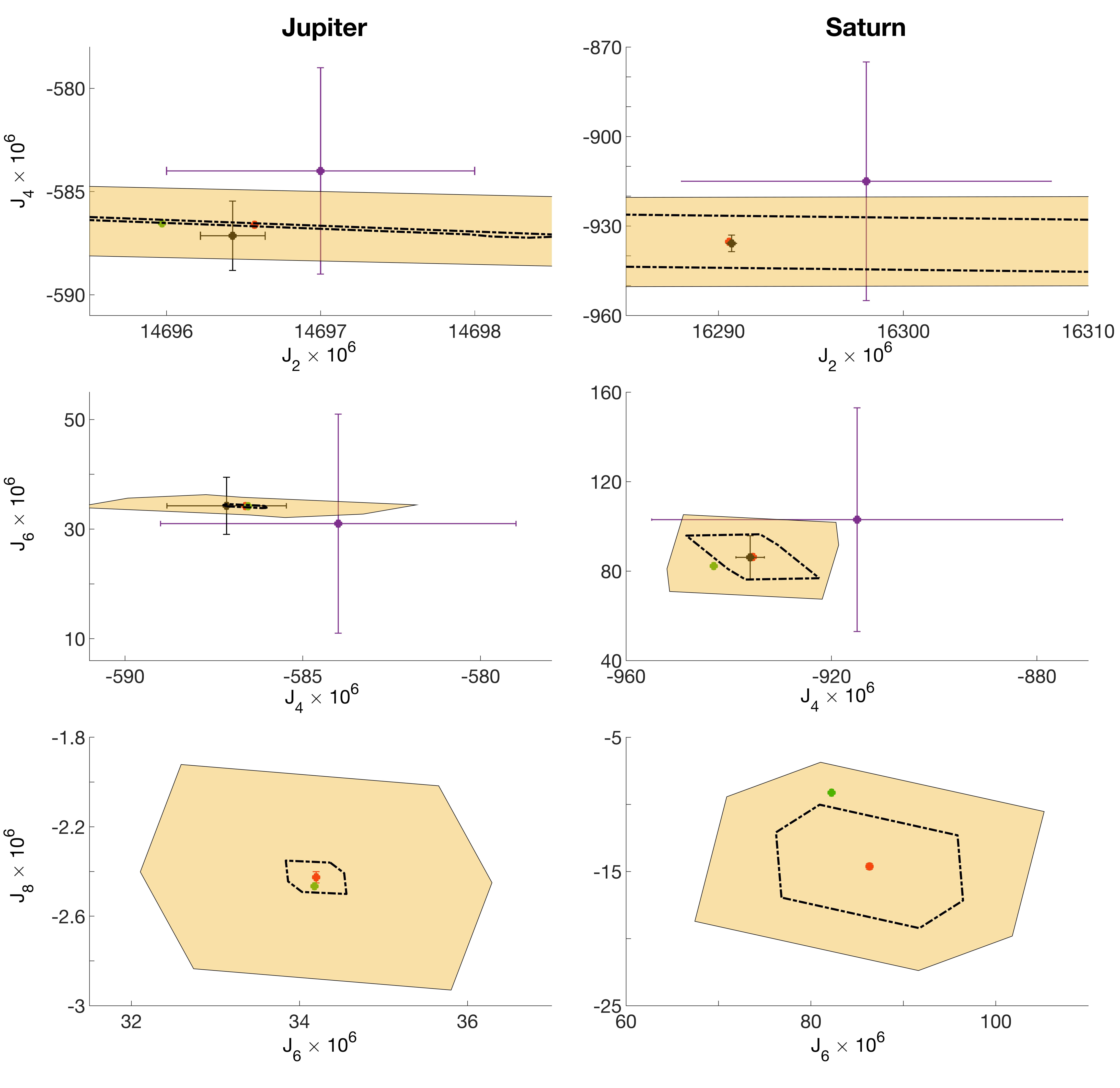}
\par\end{centering}
\caption{\label{fig:evens_history} \textbf{Supplementary Figure I: }Historical
(pre Juno and Cassini) measured values of $J_{2}$, $J_{4}$ and $J_{6}$
and their uncertainty for Jupiter (top) and Saturn (bottom) and the
Juno and Cassini measurements (including $J_{8}$, red). Voyager values
(purple) are from \citet{Campbell1985} for Jupiter and \citet{Campbell1989}
for Saturn, and Cassini values (black) are from \citet{Jacobson2003}
for Jupiter and \citet{Jacobson2006} for Saturn. Recent Juno and
Cassini results are from \citet{Iess2018} for Jupiter and \citet{Iess2019}
for Saturn. The effective uncertainty due to dynamics assuming no
knowledge on the flow and taking the widest possible range of internal
flows (see \citealp{Kaspi2017} for details) centered around the Juno/Cassini-measured
values appears in yellow, with the contour for $H=3000$~km (Jupiter)
and $H=10000$~km (Saturn) in dashed. In green is the difference
between the recent Juno/Cassini measurement and the best fit flow
profile presented in this paper ($J_{n}-\Delta J_{n}$), where $\Delta J_{n}$
is taken from column 7 in Tables \ref{tab:Jupiter-values} and \ref{tab:Saturn-values}
($\Delta J_{2}$ for Saturn is out of the range presented in that
panel), giving our best estimate to what internal models (with no
dynamics) should match to.}
\end{figure*}

\end{document}